\documentclass[preprint,preprintnumbers,amsmath,amssymb]{revtex4-1}
\usepackage[dvips]{graphicx}
\usepackage{latexsym}
\usepackage{natbib}
\usepackage[usenames]{color}
\usepackage{multirow}

\definecolor{Blue}{rgb}{0,0,0.6}
\definecolor{Green}{rgb}{0,0.6,0}
\definecolor{Red}{rgb}{0.6,0,0}

\begin{document}

\title{Origins of power-law degree distribution in the heterogeneity
  of human activity in social networks}

\author{Lev Muchnik$^{1 *}$} \author{Sen Pei$^{2,3 *}$} \author{Lucas
  C. Parra$^{4 *}$} \author{Saulo D. S. Reis$^{2,5}$} \author{Jos\'e S.
  Andrade, Jr.$^5$} \author{Shlomo Havlin$^6$} \author{Hern\'an
  A. Makse$^{2,5}$}

\affiliation{
  $^1$School of Business Administration, The Hebrew
  University of Jerusalem, 91905 Israel\\
  $^2$Levich Institute and Physics Department,
  City College of New York, New York, NY 10031, USA\\
  $^3$LMIB and School of Mathematics and Systems Science,
  Beihang University, Beijing, 100191, China\\
  $^4$ Biomedical Engineering Department,
  City College of New York, New York, NY 10031, USA\\
  $^5$Departamento de F\'{i}sica, Universidade Federal do Cear\'{a},
  60451-970 Fortaleza, Cear\'{a}, Brazil\\
  $^6$Department of Physics, Bar-Ilan University,
  52900 Ramat-Gan, Israel\\ * These authors contributed equally.\\}

\begin{abstract}
The probability distribution of number of ties of an individual in a
social network follows a scale-free power-law. However, how this
distribution arises has not been conclusively demonstrated in direct
analyses of people's actions in social networks. Here, we perform a
causal inference analysis and find an underlying cause for this
phenomenon. Our analysis indicates that heavy-tailed degree
distribution is causally determined by similarly skewed distribution
of human activity.  Specifically, the degree of an individual is
entirely random - following a ``maximum entropy attachment'' model -
except for its mean value which depends deterministically on the
volume of the users' activity.  This relation cannot be explained by
interactive models, like preferential attachment, since the observed
actions are not likely to be caused by interactions with other
people.

\end{abstract}

\maketitle

\section*{Introduction}

Millions of people edit Wikipedia pages, however, in average we find
that only $5\%$ contribute to $80\%$ of their content. Such
heterogeneous level of activity is reminiscent of the well-known and
widely applicable law postulated by Pareto~\cite{Pareto1896}, which
states that $80\%$ of the effects are induced by $20\%$ of the
causes. The example of Wikipedia users reported here highlights how
heterogeneous the activity of their users are, with both, activity
as well as degree following a power-law distribution.Indeed,
heavy-tailed distributions following a power-law have been observed
in variety of social systems ever since Pareto reported his
observation of the extreme inequality of wealth distribution in
Italy back in 1896~\cite{Pareto1896}. In recent years, due to
ubiquitous computerization, networking and obsessive data
collection, reports of heavy-tailed distributions have almost become
a
routine~\cite{Huberman1999,Axtell2001,Castellano2009,Barabasi2005,Rybski2009}.
Following simple distributions such as those of wealth, and
income~\cite{Yakovenko2009}, certain structural properties of social
systems were also found to be heavy-tailed distributed. More
specifically, distribution of the number of ties of a person
(degree) has been shown to fall in this group for vast and still
growing number of social networks~\cite{Barabasi1999,caldarelli}.
Power-law degree distributions, called
scale-free~\cite{Barabasi1999}, represent one of the three general
properties of social networks (short distances and high clustering
being the other two~\cite{Watts1998}). A power-law degree
distribution is not only the least intuitive and surprising
property, but also is the most well-studied and debated feature of
networks since extensively found in the late
90s~\cite{Barabasi1999,Faloutsis1999}.

%
%

Immediately following the empirical measurements, a number of
plausible models aiming at explaining the emergence of these
distributions have been
proposed~\cite{Barabasi1999,Caldarelli2002,Vazquez2003,Bianconi2001,Newman,Mitzenmacher2004}.
Many models reproduce heterogeneous connectivity by amplifying small
differences in connectivity -- frequently stochastically emerging --
using some kind of multiplicative process or ``preferential
attachment''
~\cite{Barabasi1999,Caldarelli2002,Vazquez2003,Bianconi2001,Newman,Mitzenmacher2004,Yule,Simon}.
Other models propose different optimization strategies leading to
scale-free~\cite{Mandelbrot,DSouza2007,Kitzak2012}.  A common
attribute of all these models is that fat-tailed distributions
emerge out of some kind of interaction between the basic system's
elements. In fact, the question is not whether there exists a
mechanism that could produce scale-free networks similar to the ones
observed, but which of the many mechanisms suggested are more likely
to actually play a significant role in each network formation.

The data presented here suggests that there is a different
underlying cause for heavy-tailed degree distributions which does
not involve interactions between people. We investigate distinct
social networks focusing on the relationship between users' activity
and degree, specifically, the number of posts, messages, or actions
of a user, i.e. {\em activity} and the number of user establishing a
link with her/him, i.e. the incoming degree, or {\em degree}, for
short.  Both, degree $k$ in the social network, and the activity $A$
of a user, exhibit power-law distributions $P(k)\sim k^{-\gamma_k}$,
and $P(A)\sim A^{-\gamma_A}$, where $\gamma_k$ and $\gamma_A$ are
the scale-free degree and activity exponents, respectively.
Positively skewed distributions of human activity were recently
reported in~\cite{microsoft,Perra2012} and we extend this result
here for a number of datasets.  More importantly, in all instances
we find that activity causally determine degree of the same user,
suggesting that the broad distribution of one, could result from the
broad distribution of the other.  It is important to note that the
studied actions are not likely to be driven by interaction with
other people.  Activity and degree, as measured here, are taken from
two different networks developed by the same pool of users, and so
there is no reason to expect that they should depend on each other
in some trivial fashion. Surprisingly, however, the number of
potential followers of a user (degree distribution) appears to be
entirely random except for its mean value, which is tightly
controlled by the volume of activity of that user. Our observations
convincingly point at the intrinsic activity of people as the
driving force behind the evolution of the examined social systems
and particularly the heterogeneity in user connectivity.  The
observed degree distribution in social systems may merely be a
manifestation of the similarly wide distribution of human activity
related to the system construction. These wide distributions in
social collaborative networks cannot be explained by interactive
model since the observed actions are not likely to be caused by
actions of other people.

\section*{Results}

\subsection*{Network construction}

We have analyzed activity of individuals over time collaboratively
working on construction of extensive electronic data sets: Wikipedia
in four different languages ({\it http://www.wikipedia.org}), and a
collaborative news-sharing web-site ({\it http://www.news2.ru}).
These datasets represent various domains of human activity and
contain records of a vast number of individual user contributions to
the collaboratively generated content (see Method). For each person,
we analyze two properties defined in two independent layers:
activity and degree. For instance, in Wikipedia, the activity
performed by users includes posting of new material and discussions
about them. This is the activity layer. Simultaneously, by tracing
users contributing to other users' personal or talk pages, we
recover the underlying network of Wikipedia contributors' personal
communication or social network. The resulting network reliably
represents actual interactions of Wikipedia
users~\cite{Kumar2004,Hsu2007,Kleinberg} and thus defines the social
network layer. The number of incoming connections, i.e. others
reaching out to the user in this network represents the degree. In
principle, activity and degree as defined here are unrelated.
Similarly, news2.ru posses the same two-layer structure of activity
and degree (see Method).


\subsection*{Analysis of activity and degree distribution}

\begin{figure}[ht]
\includegraphics[bb=0 0 978 783,width=180mm]{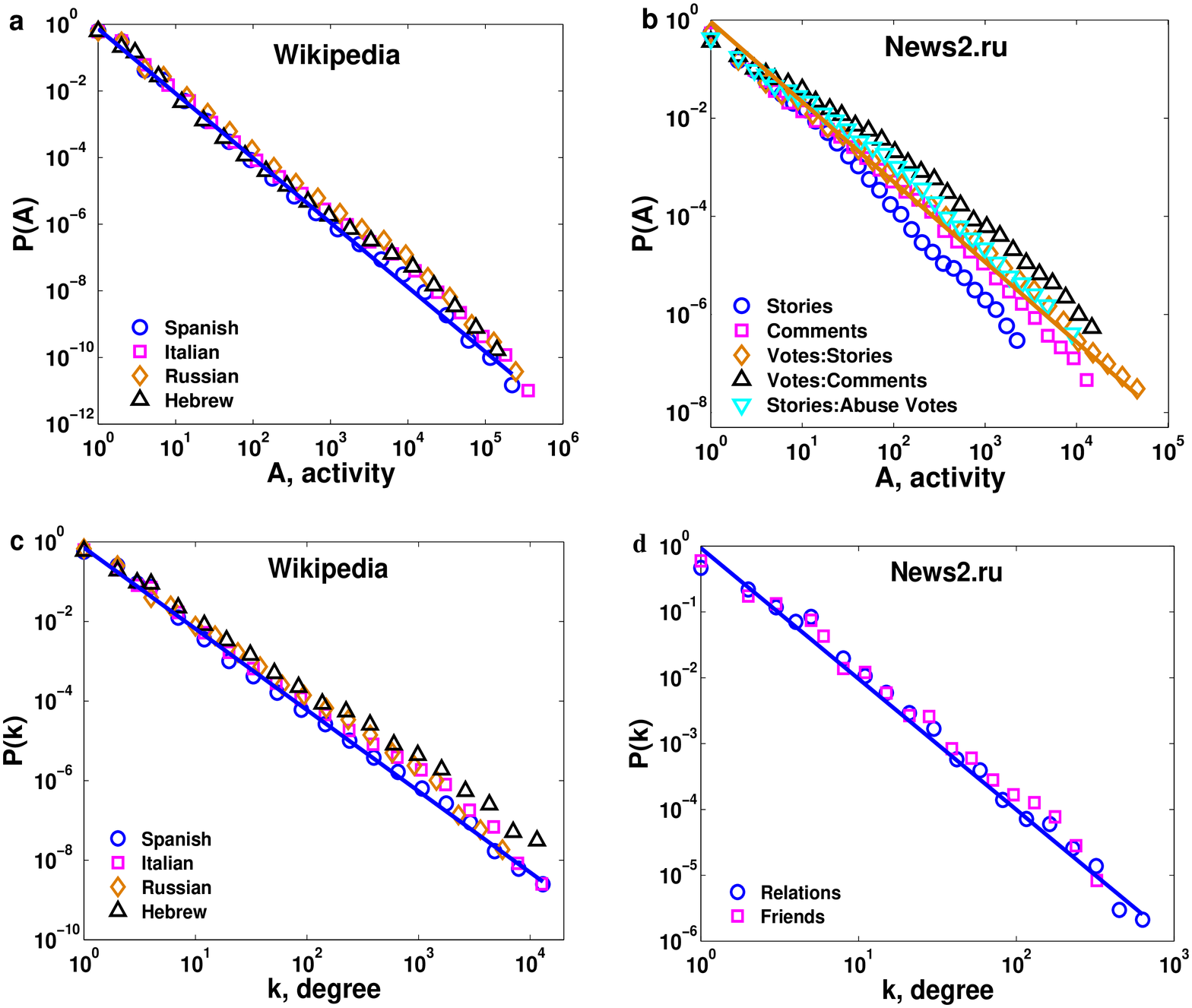}
\caption{} \label{fig1}
\end{figure}

We start by analyzing the distributions of various types of
activities performed by users in these systems. Very few of the most
active users perform the vast majority of work so that the activity
levels frequently span five orders of magnitude
(Fig.~\ref{fig1}a,b). For instance, when analyzing the activity to a
given Wikipedia page, only 5\% of users contribute $80\%$ of the
edits (Fig.~\ref{fig5} in Method). This surprising result is similar
to the $80-20$ rule postulated by Pareto~\cite{Pareto1896} to
describe the unequal distribution of wealth. Indeed,
a power-law faithfully characterizes the activity distributions in
Fig.~\ref{fig1}. The exponent of the activity distribution for
Spanish language Wikipedia is $\gamma_A = 1.752\pm 0.005$
(Fig.~\ref{fig1}a), while the activity distribution for voting in
stories in News2.ru is $\gamma_A = 1.88\pm 0.04$ (Fig.~\ref{fig1}b,
detailed fitting procedure in Method~\cite{Clauset2009,Gallos2012})
%

The activity distributions in Fig.~\ref{fig1}a~represent the number of
users as a function of the number of Wikipedia edits in four
languages. Interestingly, different populations performing similar
activity in separate instances of similarly-built social systems
exhibit identical activity distributions. Figure~\ref{fig1}b shows
several different activities performed by the same population of users
at the social news aggregator news2.ru. These activities differ in
their complexity. We consider submission of posts to be the most
difficult and time consuming of the four activity types because it
typically requires the user to locate the content on-line, evaluate
its quality and publish at the news2.ru web site by filling a form
with multiple fields. Considering the task complexity, writing
comments is arguably easier task than posting.  There are on average
nearly three comments per every published post.  These two
content-generating tasks are followed by ranking of posts and
comments. The differences in the underlying complexity of the task
seem to explain the difference in the range and slope of the observed
distributions plotted in Fig.~\ref{fig1}b.

We further observe the social networks emerging in each of these
systems. These networks serve different functions. In Wikipedia they
arise due to the direct interaction required to coordinate common
tasks. In particular, we derive social networks from the record of
edits of personal user pages by other users - a common way of
personal communication in Wikipedia (the web site rules forbid
activity-related confidential communication between its editors). In
news2.ru the social network emerges through declaration of personal
attitudes - a user may indicate that he/she likes, dislikes or is
neutral to any other user. Another social network arises from a set
of explicit (directed) declarations of friendship between news2.ru
users.
Figure~\ref{fig1} c and d present the degree distributions in these
networks. Broad distributions are measured and present in each
system, suggesting a scale-free behavior in their degree
distribution. The exponent of the degree distribution for Spanish
Wikipedia is $\gamma_k = 1.92\pm 0.01$ (Fig.~\ref{fig1}c), and for
the degree distribution in News2.ru is $\gamma_k = 2.11\pm 0.08$
(Fig.~\ref{fig1}d).

\subsection*{Dependence between activity and degree}

The present data suggest a simple explanation of the origin of
degree distributions. We first observe that the number of incoming
links aggregated by a person in all these social networks is highly
correlated to the individual's activity.  The correlation between
the degree and the activity measurements is presented in
Table~\ref{table1}. It is measured here as the correlation of the
log-values to capture the gross relationship of these two variables
across different orders of magnitude. More importantly, the
dependence analysis below suggests that the broad distribution of
activity is the driving force of scale-free degree as will be
discussed next.

It is important to emphasize that in order to avoid direct and
rather obvious correlation between different aspects of activity of
the same person, we test the correlation of individual's activity to
her degree determined by actions of his/her followers rather than
his/her own. It is possible that these actions are driven by
reciprocity, i.e., a person is simultaneously active in the
community and in constructing her social network inspiring others to
link back to her.

\begin{figure}[ht]
\includegraphics[bb=0 0 960 1013,width=180mm]{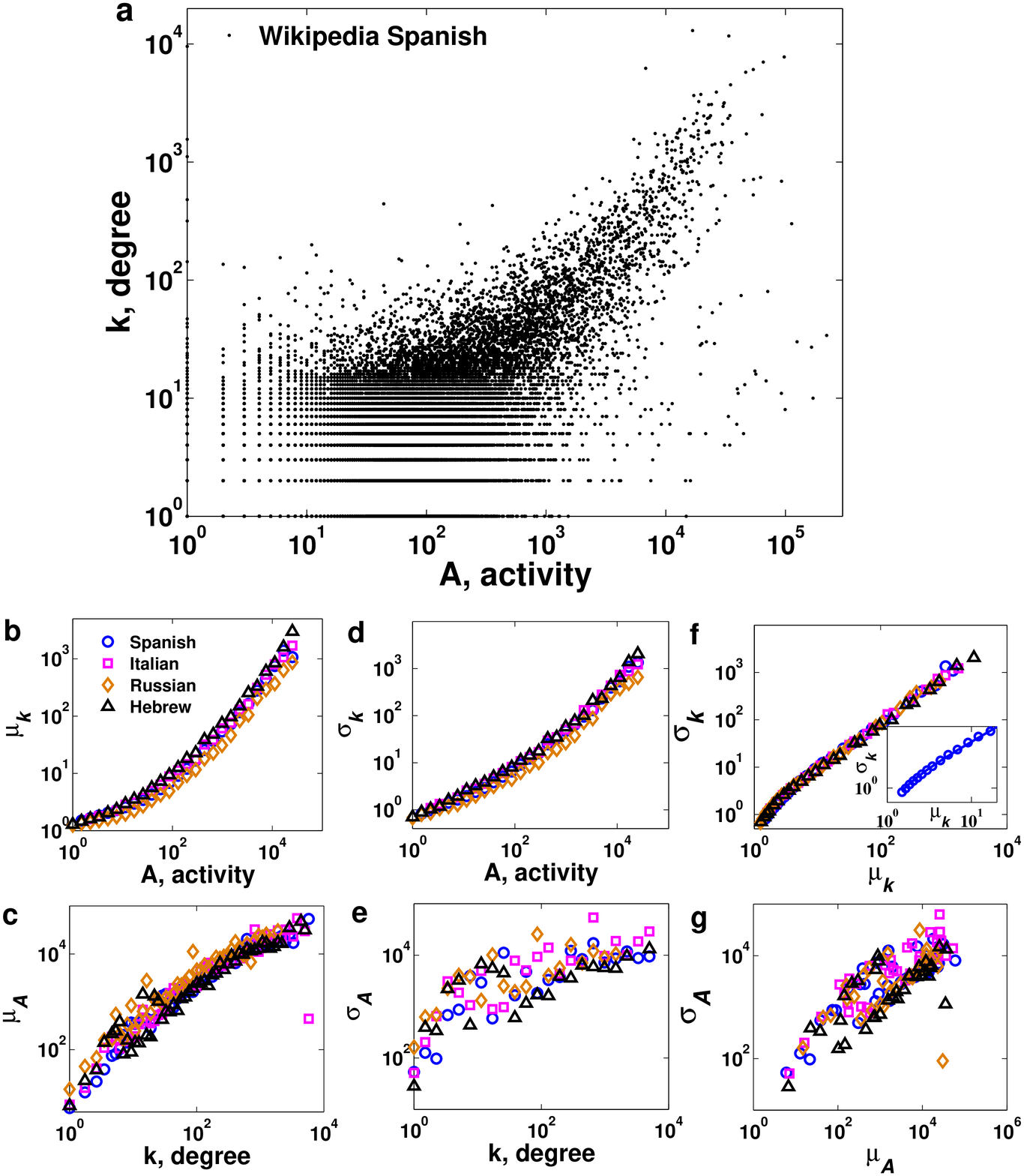}
\caption{} \label{fig2}
\end{figure}

To determine the precise nature of the $(k,A)$ relationship, we
analyze the joint distribution of degree and activity, $p(k,A)$
(Fig.~\ref{fig2}a). We find that the mean degree $\mu_k$ for a given
level of activity follows a smooth monotonic function of $A$
(Fig.~\ref{fig2}b), whereas the opposite is not true, i.e., the mean
activity $\mu_A$ does not seem to be tightly determined by degree
(Fig.~\ref{fig2}c). A similarly tight relationship exists for the
standard deviation of the degree distribution $\sigma_k$ for
specific values of the activity (Fig.~\ref{fig2}d), but, again, the
reverse is not true (Fig.~\ref{fig2}e). The conditional mean and
standard deviation of degree (conditioned on activity) show a tight
relationship with approximately unit slope $\sigma_k\approx\mu_k$
(Fig.~\ref{fig2}f). However, the $\sigma_A$, $\mu_A$ values
conditioned on degree are more variable (Fig.~\ref{fig2}g). Based on
these observations we hypothesize that the conditional degree
distribution $p(k|A)$ may be scale invariant with scale $\mu_k$
entirely determined by activity: $\mu_k=f(A)$.  Here, this
functional dependence of scale can be estimated as the mean activity
for a given $A$: $\mu_k=f(A) \approx \mathrm{mean}(k|A)$.  Indeed,
we observe that the conditional degree distribution appears to
follow a geometric distribution for all $\mu_k$:

\begin{equation}
p(k|\mu_k)=(\mu_k-1)^{(k-1)}\mu_k^{-k}. \label{geo}
\end{equation}

\begin{figure}[ht]
\includegraphics[width=88mm]{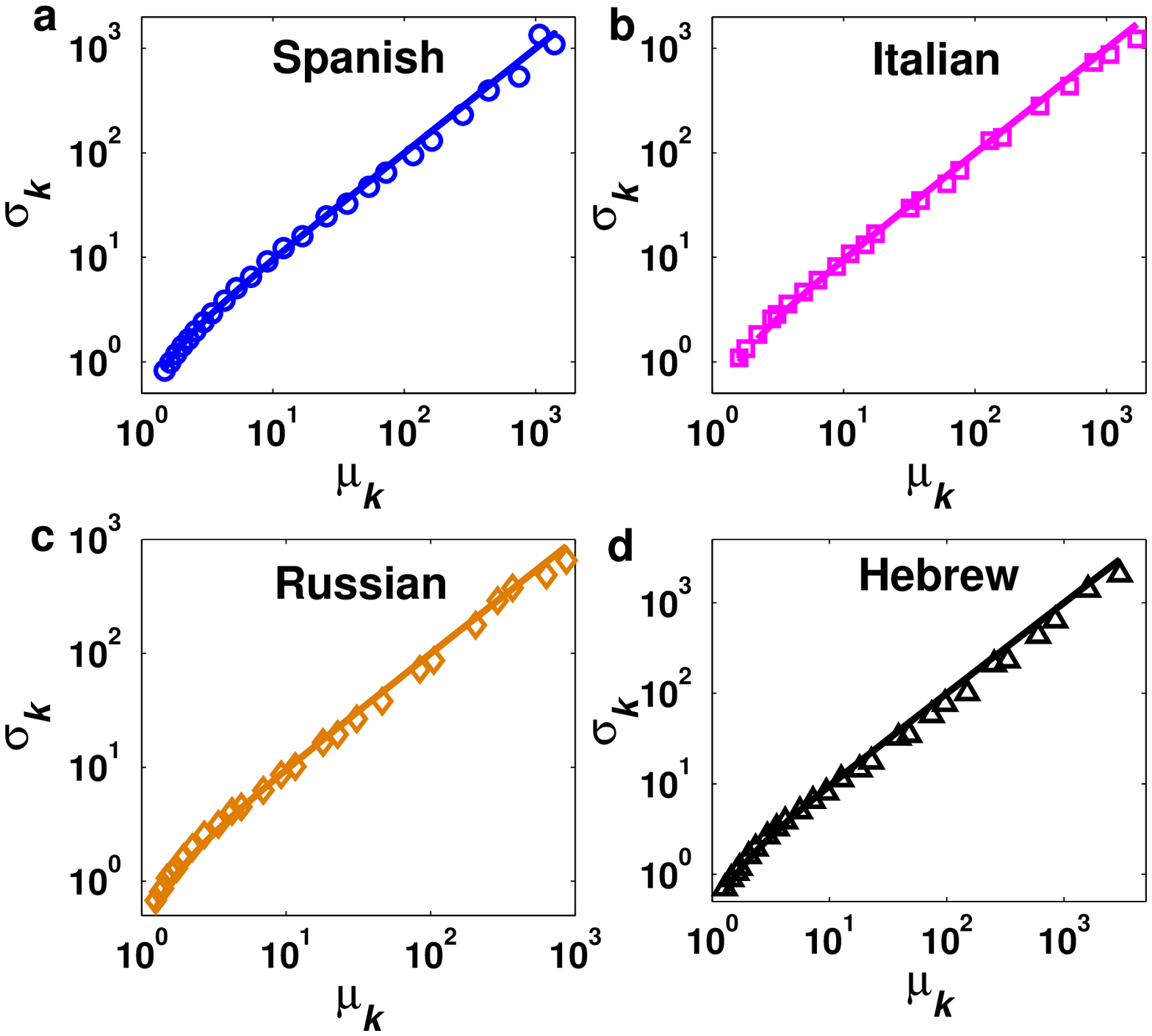}
\caption{} \label{fig3}
\end{figure}

This theoretical distribution provides a remarkably accurate fit to
the first two sample moments of degree for a given level of activity
as shown in Fig. \ref{fig3}.
We plot the standard deviation $\sigma_k$ versus mean degree $\mu_k$
for given activity for four Wikipedia databases.
The curves follow a smooth, monotonically increasing functional form
which is almost identical for all datasets (as one would expect for
activity conditioning degree). When the analysis is repeated for
activity conditioned on degree the variables do not appear to follow
a tight relationship.

The tight relationship between $\sigma_k$ versus $\mu_k$ conditioned
on activity follows asymptotically a straight line with unit slope,
which follows exactly the geometric distribution Eq. (\ref{geo}). In
Fig.~\ref{fig3}, we compare the data to the analytic relationship
between mean and standard deviation for geometric distribution Eq.
(\ref{geo}): $\mu=\frac{1}{p}$ and $\sigma=\sqrt{\frac{1-p}{p^2}}$,
where $p$ is the parameter of geometric distribution. The data fit
this theoretical curve surprisingly well for the four displayed
languages of Wikipedia ($r^2=0.8889$ in average).

\subsection*{Dependence Hypotheses}

\begin{figure}[ht]
\includegraphics[width=180mm]{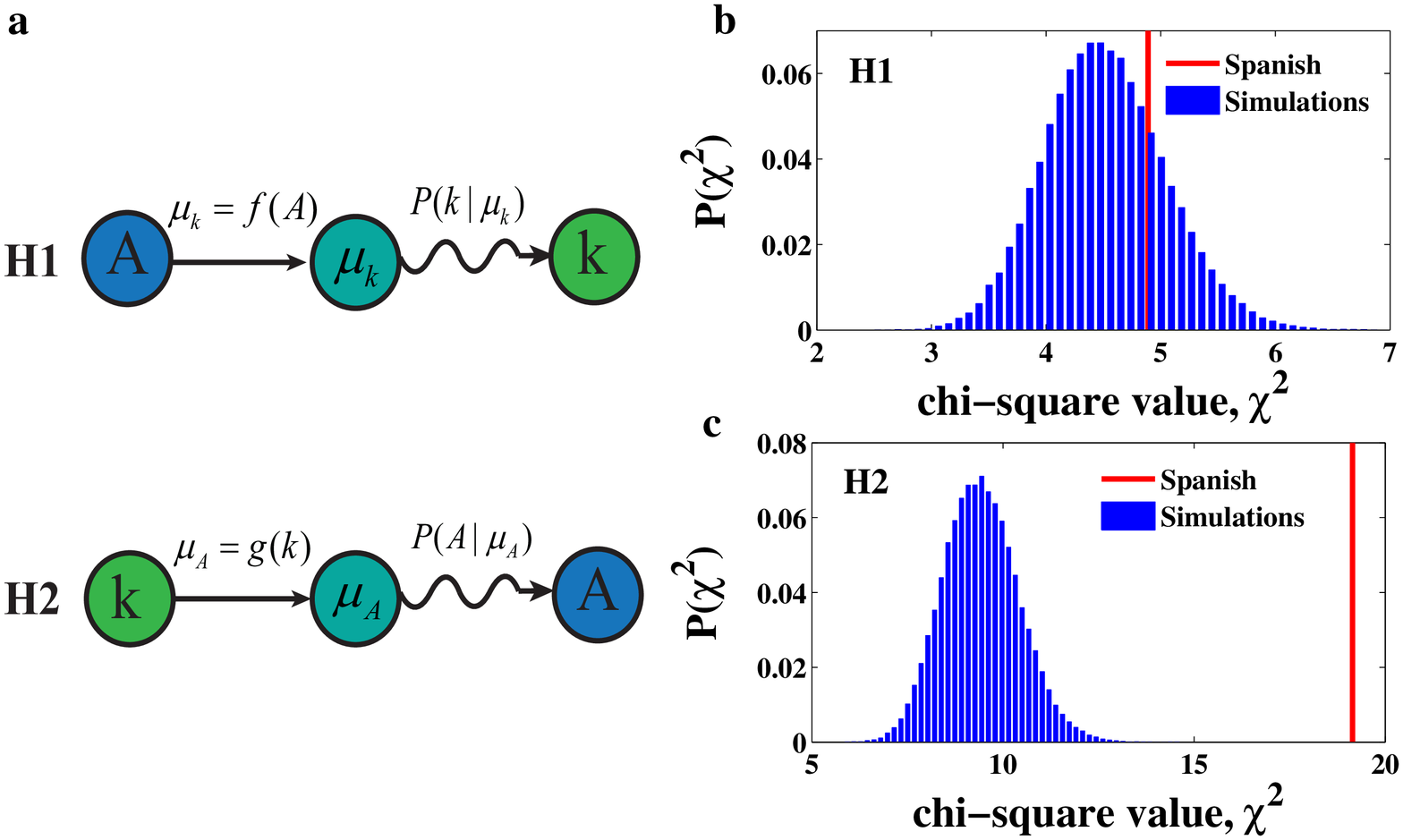}
\caption{} \label{fig4}
\end{figure}

\begin{figure}[ht]
\includegraphics[width=88mm]{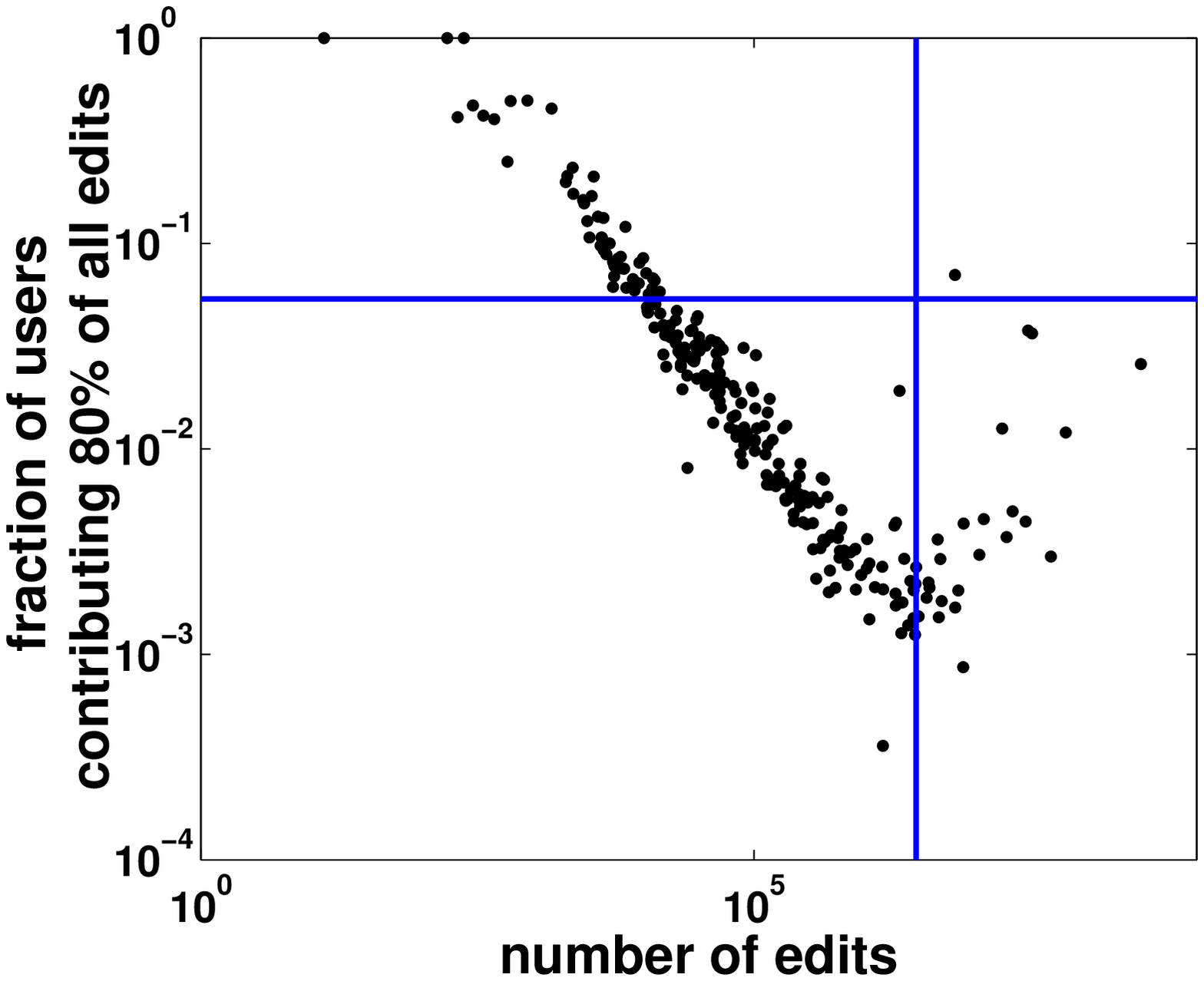}
\caption{} \label{fig5}
\end{figure}

The previous findings can be understood with the following hypothesis
H1: $A\to k$, activity deterministically affects the mean degree, but
degree is otherwise random (Fig.~\ref{fig4}a).
Note that for positive discrete variables -- like the degree -- with a
given mean, the highest entropy or least informative and most random
distribution is achieved by the geometric distribution as we find
above ~\cite{Topsoe1975}. The geometric distribution is analogous to
exponential distribution in statistical mechanics, which maximizes
entropy for continuum variables with fix mean. We also tested the
inverse hypothesis H2: $k \to A$, degree deterministically affects mean
activity, $\mu_A=g(A)\approx\mathrm{mean}(A|k)$, and activity is
otherwise random.

%
%

The goodness-of-fit of these two analytic models to histograms of H1:
activity $\to$ degree or H2: degree $\to$ activity was measured with
the $\chi$-square statistics averaged over activity or degree
respectively. The likelihood that the observed distributions match H1
or H2 was assessed using surrogate data generated with Monte-Carlo
sampling to estimate the chance occurrence of these averaged
$\chi$-square values. The results for the Spanish language Wikipedia
data indicate that we cannot dismiss the correctness of H1
(Fig.~\ref{fig4}b) with a confidence of higher than 95\% ($p=0.23$)
but that H2 can be soundly dismissed (the chance of the corresponding
$\chi$-square value occurring at random is $p<10^{-5}$). The same is
true for all other datasets (see Table~\ref{table1}). In all datasets
the likelihood of H1 is several orders of magnitudes larger than H2
and thus we accept model H1, which states that activity determines
degree.

Given the explicit model of a geometric distribution for $P(k|A)$ of
hypothesis H1, and the observed power-law distribution for activity,
$P(A) \sim A^{-\gamma_A}$, one can explicitly derive the expected
degree distribution.The conditional degree distribution closely
matches a geometric distribution (Fig.~\ref{fig3}). For large mean
values, say $\mu_k>10$, it can be very well approximated by its
continuous equivalent, the exponential distribution i.e.
$P(k|A)=\frac{1}{\mu_k}e^{-\frac{k}{\mu_k}}$. Therefore:

\begin{eqnarray}
P(k) &=& \int dAP(k|A)P(A)\sim \int
dA\frac{1}{\mu_k}e^{-\frac{k}{\mu_k}}A^{-\gamma_A}\\ &=&\int dA
A^{-\delta}e^{-\frac{k}{A^{\delta}}}A^{-\gamma_A},
(u=\frac{k}{A^{\delta}}),\\
&=&-\int\frac{du}{\delta}u^{\frac{\gamma_A-1}{\delta}}e^{-u}k^{\frac{1-\gamma_A}{\delta}-1}\\
&\sim& k^{\frac{1-\gamma_A}{\delta}-1}\sim k^{-\gamma_k}.
\end{eqnarray}

Thus the exponent is predicted to be
\begin{equation}
\gamma_{k}=1+\frac{\gamma_A-1}{\delta}.
\end{equation}
where $\delta$ defines $\mu_k \sim A^\delta$ for large $A$ as shown
in Figure~\ref{fig2}b.  The observed exponents $\gamma_k$ closely
follow these predicted exponents for all datasets
(Table~\ref{table1}).

\section*{Discussion}

The causal inference argument provided here is borrowed from ideas
recently developed in causal
inference~\cite{Pearl2009,Hoyer2009,Zhang2009}. There, a deterministic
functional dependence of cause on mean effect is hypothesized and
deviations from this mean effect are assumed to have fixed standard
deviation but to be otherwise random. With two variables for which one
wishes to establish causal direction, the model is evaluated in both
directions and the more likely one is postulated to indicate the
correct causal dependence, as we have done here. This approach has
been demonstrated to give the correct causal dependence for a large
number of known causal relationships~\cite{Janzing2012}, and
theoretical results indicate that there is only an exceedingly small
class of functional relationships and distributions for which this
procedure would give the incorrect answer. Such an identifiability
proof does not yet exist for the present case where the standard
deviation is not constant. Nevertheless, our explicit model of a
deterministic effect of human activity on the success of establishing
social links is the simplest possible explanation for the data
available to us. For a different dataset a different probabilistic
model may be better suited.

The individual activity of people deterministically affects the mean
success at establishing links in a social network, and the specific
degree of a given user is otherwise random following a maximum entropy
attachment (MEA) model. The MEA model is exemplified in
Fig. \ref{fig4}a and consists of the following steps: Introduce a node
$i$ with $q$ links, where $q$ is drawn from a probability given by the
activity of the node. The activity has an intrinsic power-law
distribution. Then, link the $q$ links at random following maximum
entropy principle with the concomitant geometric distribution
$P(k|\mu_k)$. This mechanism contrasts with the preferential
attachment mechanism
\cite{Barabasi1999,Newman,Mitzenmacher2004,Yule,Simon} where each link
attaches to a node with a probability proportional to the number of
links of that node.  A possible mechanism by which a geometric
distribution could arise is based on the notion of ``success''. In
this model, the activity of users aims to achieve a specific outcome
(a Wikipedia project), and each new incoming link can aid in achieving
this desired outcome; once the goal is achieved the user stops
collecting links. The probability of the desired event in this model
is $q=1/\mu_k \sim A^{-\delta}$. Hence, those users working so very
hard may have an exceedingly unlikely event they are aiming for. But
eventually, they too will succeed, and will turn their attention away
from the on-line social network.

The present data indicates that degree distribution is maximally
random except for what can be determined solely from the volume of a
user's activity. Does this mean that the precise content of a user's
actions (the meaning and quality of the edits in Wikipedia, messages,
etc) is immaterial in determining his/her success in establishing
relationships? One can only hope that small deviations from this
maximum entropy attachment model will become more pronounced with
increasing data-set sizes, which can then point us to the benefits of
well thought out and carefully executed actions, specially in
specialized large-scale collaborative projects like Wikipedia.

Whether the dynamics of preferential attachment is consistent with the
maximum entropy distribution of degree remains to be established.
What is certain is that distributions of levels of activities in all
tested populations are heavily heavy-tailed indicating highly varying
level of involvement of users in collaborative efforts. We showed here
that this fact alone is sufficient to produce the heavy-tailed
distribution of degree observed throughout social networks.
Therefore, previous interactive models may not be necessary. The
present result shifts the burden of proof to explaining the origin to
the incredible diversity in human effort observed here spanning five
orders of magnitude.

\section*{Method}

\subsection*{Datasets} \label{datasets}

The number of actions contained in the datasets range from hundreds
of thousands to hundreds of millions of user actions. From the
editing on Wikipedia, to the votes, to commentaries on News2.ru,
these actions represents different and natural underlying dynamics
of social networks, since they range from collaborative interaction
(Wikipedia) to discussions about different interesting of human
behavior (New2.ru), which are intrinsic properties of the social
nature of the web.

We have collected details about user activity in the Wikipedia
project and reconstructed the underlying social network. In addition
to the widely used term and category pages, Wikipedia provides
special pages associated with specific contributing authors and
discussion (talk) pages maintained alongside each of these pages.
These user pages are widely used by Wikipedia contributors for
coordination behind the scenes of the project. In fact, interaction
via user and discussion pages dominates all other communication
methods. However, communication via personal user pages (and the
corresponding discussion pages) differs from the topic-associated
talk pages in that it is explicit person-to-person communication
rather than general topic specific, usually impersonal
communication. By tracing users contributing to other user's
personal or talk pages, we recover the underlying network of
Wikipedia contributor's personal communication. Not surprisingly, as
presented in the next section, the obtained social networks show a
scale-free degree distribution, typically observed in a variety of
social networks analyzed so far.

The other data set is a de-identified record of activities of social
news aggregator news2.ru. The record contains all actions performed
by the community members over more than three years of collaborative
selection and discussion of news-related content. These,
user-related actions include such events as submission of news
article, comments as well as preference-revealing actions such as
voting for articles (``dig'' or ``bury'', using digg.com language)
and other users' comments. In addition to the trace of user
activity, the data contains explicit social network layer. Each user
may publicly declare his/her (positive, neutral or negative)
attitude to any other user. Considering the personal flavor of the
rather emotional way people interact through commentary threads,
this list of attitudes when aggregated can be perceived as social
network. In addition, users maintain list of friends, usually
including users most favorable on them. These networks are
directional, which allows to focus on the incoming links, since they
can not be controlled by the target individual, but by his/her
friends.

Each of these systems represents different approaches to
collaborative content creation. The Wikipedia editors interact to
create the same content collaboratively so that the content
contributed by one user can be complemented, altered or completely
removed by others. The news2.ru represents a mixed case in which the
content is contributed individually, but collaboratively ranked.
Given these fundamental differences in user activity and network
dynamics, the similarities between these systems reported below are
particularly revealing.

\subsection*{Method of Power-law Fitting} \label{maxlike}

To get the exponents $\gamma_k$ and $\gamma_A$ of power-law
distribution, we present a rigorous statistical test based on
maximum likelihood methods~\cite{Clauset2009}. Take the degree
distribution as an example. We fit degree distribution assuming a
power law within a given interval. For this, we use a generalized
power-law form
\begin{equation}
P(k; k_{min},
k_{max})=\frac{k^{-\gamma}}{\zeta(\gamma,k_{min})-\zeta(\gamma,k_{max})},
\end{equation}
where $k_{min}$ and $k_{max}$ are the boundaries of the fitting
interval, and the Hurwitz $\zeta$ function is given by
$\zeta(\gamma,\alpha)=\sum_{i}(\gamma+\alpha)^{-\gamma}$.

We use the maximum likelihood method, following the rigorous
analysis of Clauset et al.~\cite{Clauset2009}. The fit was done in
an interval where the lower boundary was $k_{min}$. For each
$k_{min}$ value we fix the upper boundary to $k_{max}=K$, where $K$
is the maximal degree. We calculate the slopes in successive
intervals by continuously increasing $k_{min}$ and varying the value
of $w$. In this way, we sample a large number of possible intervals.
For each one of them, we calculate the maximum likelihood estimator
through the numerical solution of
\begin{equation}
\gamma=argmax(-\gamma\sum_{i=1}^{N}\ln
k_i-N\ln[\zeta(\gamma,k_{min})-\zeta(\gamma,k_{max})]),
\end{equation}
where $k_i$ are all the degrees that fall within the fitting
interval, and $N$ is the total number of nodes with degrees in this
interval. The optimum interval was determined through the
Kolmogorov-Smirnov (KS) test.

For the goodness-of-fit test, we use the Monte Carlo method
described in ~\cite{Clauset2009}. For each possible fitting
interval, we calculate the Kolmogorov-Smirnov statistics $D$ for the
obtained cumulative distribution function. Then we choose the
interval with the minimal $D$ as the best fitting interval and take
the $\gamma$ in this interval as the final result. As to the
standard error estimation, we adopt the method in
~\cite{Clauset2009}. The standard error on $\gamma$, which is
derived from the width of the likelihood maximum, is
$e=(\gamma-1)/\sqrt{n}+O(1/n)$, where $n$ is the number of data.

Although the fitting method mentioned above is rigorous, it is
suitable for fitting probability density distributions. When we fit
the data $\mu_k=A^{\gamma_A}$, we use another fitting
method~\cite{Gallos2012}. The procedure for determining fitting
interval is similar. In each fitting intervals, the fittings were
done using ordinary least squares methods. The goodness of fitting
was estimated through the coefficient of determination, $r^2$, where
$0\leq r^2\leq 1$. The value of $r^2$ is used as a measure of how
reliably the fitted line describes the observed points, and is often
described as the ratio of variation that can be explained by the
fitted curve over the total variation.  We assume that any value
above $r^2\geq0.85$ represents an accepted fitting. The final result
is the average of the accepted exponent.

\subsection*{Users contributing to $80\%$ of a Wikipedia page} \label{Wiki}
In Fig.\ref{fig5}, each dot represents a distinct Wikipedia project
page. Horizontal axis measures the total number of edits for each
project. Vertical axis represents the fraction of contributors to
that project who performed $80\%$ of edits on that project. This
fraction drops fast (with power law) as the number of edits grows.
This suggests that the largest projects are dominated by a few very
dedicated users. Perhaps more representative are the mean values;
the vertical line indicates the average edits and the horizontal
line marks the fraction of users contributing $80\%$ if the work in
the average across projects (approximately $5\%$)

\subsection*{Monte-Carlo sampling for hypothesis tests} \label{surrogate}
The accuracy of fit of the data to the theoretical geometric
distribution is measured as the $\chi^2$ goodness-of-fit to the
conditional histogram. As an example, consider H1 for the Spanish
Wikipedia data: For the theoretical distribution we use for each
activity the mean degree $\mu_k$ as shown in Fig.~\ref{fig2}b. The
$\chi^2$ value is then averaged over all activity bins shown in that
figure. To test if this observed average $\chi^2$ is consistent with
chance assuming H1 we generate surrogate data following H1: For each
given activity, we generate the same amount of random numbers from a
geometric distribution with the same mean values, calculate the
$\chi^2$ values and again, average across activities. We draw $10^5$
such samples and obtain a distribution of average $\chi^2$
(Fig.~\ref{fig4}b). The chance that the $\chi^2$ for the Spanish
Wikipedia data occurred by chance (p-value) is the fraction of times
the surrogate data provided a value larger than the one observed
(red line in Fig.\ref{fig4}b). The analysis for H2 is analogous
using the data as shown in Fig.~\ref{fig2}c. The resulting p-values
for all datasets can be found in Table I.

\section*{Acknowledgments}

We thank G. Khazankin, Research Institute of Physiology SB RAMS for
kindly providing access to invaluable data on news2.ru user
activity. The research is supported by NSF Emerging Frontiers, ARL,
FP7 project SOCIONICAL and MULTIPLEX, CNPq, CAPES, and FUNCAP.

\section*{Author contributions}
H.A.M., S.H. and J.S.A. designed research. L.M. prepared data. L.M.,
S.P., L.C.P. and S.D.S.R. analyzed the data. All authors wrote,
reviewed and approved the manuscript.

\section*{Additional information}
Competing financial interests: The authors declare no competing
financial interests.

\newpage


%
\begin{table*}[ht]
\begin{tabular}{c|c|c|c|c|c|c|c}
\hline\hline
Networks & $r_{\log}$ & $p_{\rm H1}$ & $p_{\rm H2}$ & $\delta$ & $\gamma_A$ & $\gamma_k$ & predicted $\gamma_k$\\
\hline
Spanish & $0.64$ & $0.23$ & $<10^{-5}$  & $0.79\pm 0.02$ & $1.752\pm 0.005$ & $1.92\pm 0.01$ & $1.95\pm 0.03$ \\
Italian & $0.69$ & $0.11$ & $<10^{-5}$  & $0.70\pm 0.04$ & $1.620\pm 0.004$ & $1.85\pm 0.01$ & $1.88\pm 0.05$\\
Russian  & $0.69$ & $0.13$ & $<10^{-5}$ & $0.68\pm 0.03$ & $1.618\pm 0.007$ & $1.89\pm 0.01$ & $1.91\pm 0.05$\\
Hebrew  & $0.77$ & $0.16$ & $<10^{-5}$  & $0.67\pm 0.04$ & $1.574\pm 0.008$ & $1.80\pm 0.01$ & $1.85\pm 0.05$\\
\hline
Story  & $0.64$ & $0.10$ & $<10^{-5}$ & $0.79\pm 0.08$ & $1.98\pm 0.04$ & $2.11\pm 0.08$ & $2.2\pm 0.1$\\
Comment & $0.68$ & $0.37$ & $<10^{-5}$  & $0.72\pm 0.09$ & $1.88\pm 0.05$ & $2.11\pm 0.08$ & $2.2\pm 0.2$ \\
Story Vote & $0.65$ & $0.05$ & $<10^{-5}$ & $0.70\pm 0.08$ & $1.88\pm 0.04$ & $2.10\pm 0.08$ & $2.3\pm 0.2$\\
Comment Vote & $0.59$ & $0.26$  & $<10^{-5}$ & $0.71\pm 0.09$ & $1.85\pm 0.09$ & $2.1\pm 0.2$ & $2.2\pm 0.2$\\
\hline\hline
\end{tabular}
\caption{ Statistics for different datasets. The log-correlation
  $r_{\log}$ between the user's activity and his/her degrees in
  Wikipedia and News2.ru is displayed in the first column. $p_{\rm
    H1}$ and $p_{\rm H2}$ are p-values for hypotheses H1 and H2,
  respectively. $\delta$ is the exponent for $\mu_k \sim A^{\delta}$,
  while $\gamma_A$ and $\gamma_{k}$ are the power law exponents of
  activity and degree distribution obtained by fitting the data. The
  predicted $\gamma_k$ results from scaling relation as detailed in
  the text.}
\label{table1}
\end{table*}
%

\newpage

Fig. 1. Probability distribution of activities and degree. (a)
Probability density function of Wikipedia contributors as a function
of the number of performed page edits in four languages. (b)
Probability density function of news2.ru for five different
activities.Lines indicate power-law fitting for Spanish and Stories
with the maximum likelihood methods. (c) Probability distribution of
degree for social networks as a function of number of links between
Wikipedia contributors. Degree represents the number of links other
users establish with a given user. (d) Distribution for networks of
relationship (positive/negative) between users of news2.ru web
portal and users' friendships.

Fig. 2. Analysis of joint distribution of activity and degree. (a)
Scatter plot of degree and activity for each user in Wikipedia
Spanish dataset. (b) Mean degree $\mu_k$ for given activity. (c)
Mean activity $\mu_A$ for given degrees. (d) Standard deviation of
degree $\sigma_k$ for given activity. (e) $\sigma_A$ for given
degree. (f) Relationship between standard deviation of degree
$\sigma_k$ and the mean value $\mu_k$ for given activity. Inset is
the theoretical fit of geometric distributions for Spanish
Wikipedia. (g) $\sigma_A$ versus $\mu_A$ for given degree.

Fig. 3. Test of ``maximum entropy attachment model'' via the
geometric distribution. Theoretical relationship of mean and
standard deviation for geometric distribution (solid curve) and data
points for Wikipedia in four languages.

Fig. 4. Causal hypotheses and test result. (a) Schematic diagrams
for hypotheses H1 and H2. H1: Mean degree is determined by activity
through function $\mu_k=f(A)$. Then degree is random distributed
according to the conditional probability distribution $P(k|\mu_k)$.
H2 is the other way around. (b) and (c) Results of Monte-Carlo
simulation with $10^5$ samples following H1 and H2 for the Spanish
Wikipedia data.  The vertical red lines show the goodness-of-fit
$\chi^2$ of the actual data to H1 and H2, respectively. The
empirical analysis clearly favors H1 over H2.

Fig. 5 (color online). Users contributing to $80\%$ of a Wikipedia
page.

\newpage


%
%

\newpage


%
%

\newpage


%
%

\newpage

%
%

\newpage

%
%

\end{document}